\documentclass[a4paper,fleqn]{cas-dc}

\usepackage[numbers]{natbib}

\def\tsc#1{\csdef{#1}{\textsc{\lowercase{#1}}\xspace}}
\tsc{WGM}
\tsc{QE}
\tsc{EP}
\tsc{PMS}
\tsc{BEC}
\tsc{DE}

\shorttitle{Z. Zhang et al.: Learning Frame Level Attention for Environmental Sound Classification}
\shortauthors{Z. Zhang et al.}

\title [mode = title]{Learning Frame Level Attention for Environmental Sound Classification}

\author[1]{Zhichao Zhang}

\address[1]{Shanghai Institute for Advanced Communication and Data Science, Shanghai University, Shanghai, 200444, China}

\author[1]{Shugong Xu}
\cormark[1]
\ead{shugong@shu.edu.cn}

\author[1]{Shunqing Zhang}

\author[1]{Tianhao Qiao}

\author[1]{Shan Cao}

\begin{document}

\maketitle

\begin{abstract}
Environmental sound classification (ESC) is a challenging problem due to the complexity of sounds. The classification performance is heavily dependent on the effectiveness of representative features extracted from the environmental sounds. However, ESC often suffers from the semantically irrelevant frames and silent frames. In order to deal with this, we employ a frame-level attention model to focus on the semantically relevant frames and salient frames. Specifically, we first propose a convolutional recurrent neural network to learn spectro-temporal features and temporal correlations. Then, we extend our convolutional RNN model with a frame-level attention mechanism to learn discriminative feature representations for ESC. We investigated the classification performance when using different attention scaling function and applying different layers. Experiments were conducted on ESC-50 and ESC-10 datasets. Experimental results demonstrated the effectiveness of the proposed method and our method achieved the state-of-the-art or competitive classification accuracy with lower computational complexity. We also visualized our attention results and observed that the proposed attention mechanism was able to lead the network tofocus on the semantically relevant parts of environmental sounds.
\end{abstract}



\begin{keywords}
Environmental Sound Classification \sep Convolutional Recurrent Neural Network \sep Attention Mechanism
\end{keywords}

\section{Introduction}

Environmental sound classification (ESC) has received increasing research attention during recent several years, which is widely applied in surveillance \cite{radhakrishnan2005audio}, home automation \cite{vacher2007sound}, scene analysis \cite{barchiesi2015acoustic} and machine hearing \cite{lyon2010machine}. Different from music and speech recognition tasks, ESC has quite limited pre-known knowledge with respect to the temporal and frequency characteristics. In addition, as the environmental sounds usually behaves in a variable fashion, several naive deterministic prediction schemes are often failed to obtain good performance, which leads ESC to be more challenging in nowadays.

To address this challenge, a variety of signal processing and machine learning techniques have been applied for ESC. For the former, some naive features, e.g. zero-crossing rate or short-time energy, are analyzed via some heuristic backends. With the development of signal processing skills, some dictionary-based methods, such as dictionary learning \cite{chu2009environmental}, matrix factorization \cite{bisot2017feature}, are successfully applied in ESC \cite{mesaros2018detection}. However, this type of schemes often requires tedious feature design process to obtain reasonable accuracy. In addition, some machine learning techniques, including gaussian mixture model (GMM) \cite{dhanalakshmi2011classification} and support vector machine (SVM) \cite{chu2009environmental} have been widely adopted in ESC. Since these techniques have ability to handle complex high-dimensional features, multiple feature transformation schemes have been applied, such as mel-frequency cepstral coefficient (MFCC), mel-spectrogram features \cite{chu2009environmental}, gammatone-spectrogram features \cite{valero2012gammatone} and wavelet-based features \cite{geiger2015improving}.

In recent years, deep neural networks (DNNs) have shown outstanding performance in feature extraction for ESC. Compared to hand-crafted feature, DNNs have the ability to extract discriminative feature representations from large quantities of
training data and generalize well on unseen data. McLoughlin et al. \cite{mcloughlin2015robust} proposed a deep belief network
to extract high-level feature representations from magnitude spectrum, which yielded better results than the traditional methods. Piczak \cite{piczak2015environmental} first evaluated the potential of convolutional neural network (CNN) in classifying short audio clips of environmental sounds and showed excellent performance on several public datasets. In order to model the sequential dynamics of environmental sound signals, Vu et al. \cite{vu2016acoustic} applied a recurrent neural network (RNN) to learn temporal relationships.

\begin{figure*}
\centering
    \includegraphics[width=5.5 in]{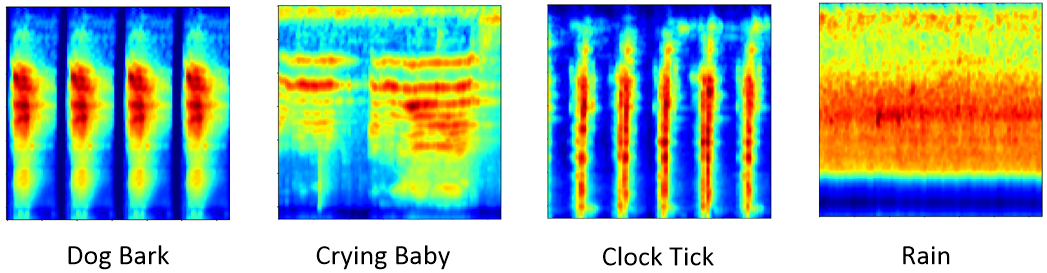}
        \caption{Examples of log gammatone spectrogram in ESC-50 dataset. From left to right, the class is \emph{dog bark}, \emph{crying baby}, \emph{clock tick} and \emph{rain}.}
        \label{fig:framelevel_att}
\end{figure*}

However, these schemes are limited to improve classification accuracy since they usually ignore the complex temporal characteristics of environmental sounds. Generally speaking, the temporal structure of environmental sounds can be transient (e.g. \emph{gun shot}), continuous (e.g. \emph{rain}) or intermittent (e.g. \emph{dog bark}), which makes it unfeasible to simply model the temporal variations via the existing techniques like hidden markov model (HMM). In addition, sound clips usually contain many periods of silence in public ESC datasets, with only a few intermittent frames associated with the characteristics of sound classes. Figure \ref{fig:framelevel_att} shows some examples of log gammatone spectrogram in ESC-50 dataset. We see that some salient and semantically relevant features only distribute in a few frames and the features usually contain silent or noisy frames, which reduces the robustness of model and increase misclassification. To deal with the problems, we explore frame-level attention mechanisms for CNN layers and RNN layers to help the network focus on semantically relevant frames. Attention mechanisms have shown outstanding performance in learning relevant feature representations for sequence data and have been successfully applied to a wide variety of tasks, including speech recognition \cite{chorowski2015attention}, machine translation \cite{bahdanau2014neural, sankaran2016temporal}, document classification \cite{yang2016hierarchical} and so on. In the field of ESC, several works \cite{guo2017attention,WJ2018ASC,ZR2018ASC,li2019multi,zhang2019attention} have studied the effectiveness of attention mechanisms and have obtained promising results in several datasets.

In this paper, we proposed an attention mechanism based convolutional RNN architecture (ACRNN) in order to focus on semantically relevant frames and produce discriminative features for ESC. In our proposed technique, softmax is used as scaling function to generate attention weights \cite{zhang2019attention}. However, the softmax based attention will be forced to focus on a few frames with large weights, which is not ideal for some types of environmental sounds, especially for continuous signals. It is also shown that the softmax based attention does not work well for CNN layers. Therefore, we investigated a sigmoid based attention for CNN layers and compared the classification accuracy with the softmax based one. In addition, this type of method usually requires a large amount of training data and the current public ESC datasets are quite limited. Therefore, we applied an efficient data augmentation scheme named mixup for ESC, which was originally proposed to handle image-related tasks \cite{zhang2017mixup}. We evaluated our proposed method on the ESC-10 and ESC-50 datasets and our model achieved the state-of-the-art or competitive performance in terms of classification accuracy. Furthermore, we visualized the learned attention results in order to give a better understanding of how the proposed frame-level attention helps recognize different environmental sounds. The main contributions of this paper are summarized as follows.

\begin{itemize}
    \item  To deal with silent frames and semantically irrelevant frames, We employ an attention model to automatically focus on the semantically relevant frames and produce discriminative features for ESC. We explore both the performance of frame-level attention mechanism for CNN layers and RNN layers. We investigate the selection of attention scaling function and visualize the attention results to have a better understanding how frame-level attention works.
    \item  To analyze temporal relations, We propose a novel convolutional RNN model which first uses CNN to extract high level feature representations and then inputs the features to bidirectional GRUs for temporal summarization. We combine the convolutional RNN and attention model in a unified architecture.
    \item To further improve classification performance, we apply a data augmentation pipeline which is applied directly to audio spectrogram of ACRNN via mixup and helps the network learn useful features.
\end{itemize}

The rest of this paper is organized as follows. Recent related works of ESC are introduced in Section~\ref{sect:rela-work}. Section~\ref{sect:methods} provides detailed description about the proposed methods, including feature extraction, network architecture, frame-level attention mechanism and data augmentation. Section~\ref{sect:exp} provides the experimental settings and results on the ESC-10 and ESC-50 datasets. Finally, Section~\ref{sect:conc} concludes the paper.

\section{Related Work} \label{sect:rela-work}
In this section we introduce the recent deep learning methods for environmental sound classification.

The 2-D CNNs are originally used to analyze spectrogram-like features. Piczak \cite{piczak2015environmental} first proposed to apply a 2-D CNN to learn the log mel spectrogram features and obtained a significant improvement than KNN, SVM and random forest. Since log mel spectrogram could be considered as a 2-D image, several well-known image recognition networks, including AlexNet and GoogLeNet, have been adopted to classify the input spectrogram features for ESC \cite{boddapati2017classifying}. More recently, Zhang et al. \cite{zhang2018deep} compared the performance between log mel spectrogram and log gammatone spectrogram features for ESC and reported that the log gammatone spectrogram based system performed better in terms of classification accuracy.

Some researchers also proposed to learn the representations directly from 1-D raw waveform data. Dai et al. \cite{dai2017very} proposed to use 1-D convolution and pooling to learn classification model and showed competitive accuracy with log mel spectrogram based methods, however, required more convolutional layers (up to 34 layers). Tokozume et al. \cite{tokozume2017learning} proposed an end-to-end network named EnvNet, which first used 1-D convolution to learn 2-D feature maps that were then classified via 2-D convolution and pooling. In  \cite{aytar2016soundnet}, Aytar et al. proposed to learned rich sound features from a large amount of unlabeled videos and then adapted the learned sound features to target datasets, where the sound recognition network utilized 1-D convolution.

Recently, attention mechanisms have been incorporated to improve classification performance in the field of ESC \cite{WJ2018ASC,ZR2018ASC,li2019multi,guo2017attention,zhang2019attention}. Guo et al. \cite{guo2017attention} proposed to apply attention to a convolutional LSTM network, where the attention is calculated via a weighted sum up of the output of LSTM layers along time dimension. Closed to  \cite{guo2017attention}, Wang et al. \cite{WJ2018ASC} proposed to stack multiple attention network to get more powerful representations. In addition, Li et al. \cite{li2019multi} proposed a temporal attention mechanism for convolutional layers to enhance the representative ability of CNN, which was calculated from input spectrogram and re-weighted the CNN feature maps via dot-product operation along the time dimension. In this paper, we investigate the frame-level attention mechanism for CNN layers and RNN layers and provide a visualization results of our attention to give a better understanding how frame-level attention helps recognize different environmental sounds.

\section{Methods} \label{sect:methods}

\begin{figure*}
\centering
        \includegraphics[width=6.8 in]{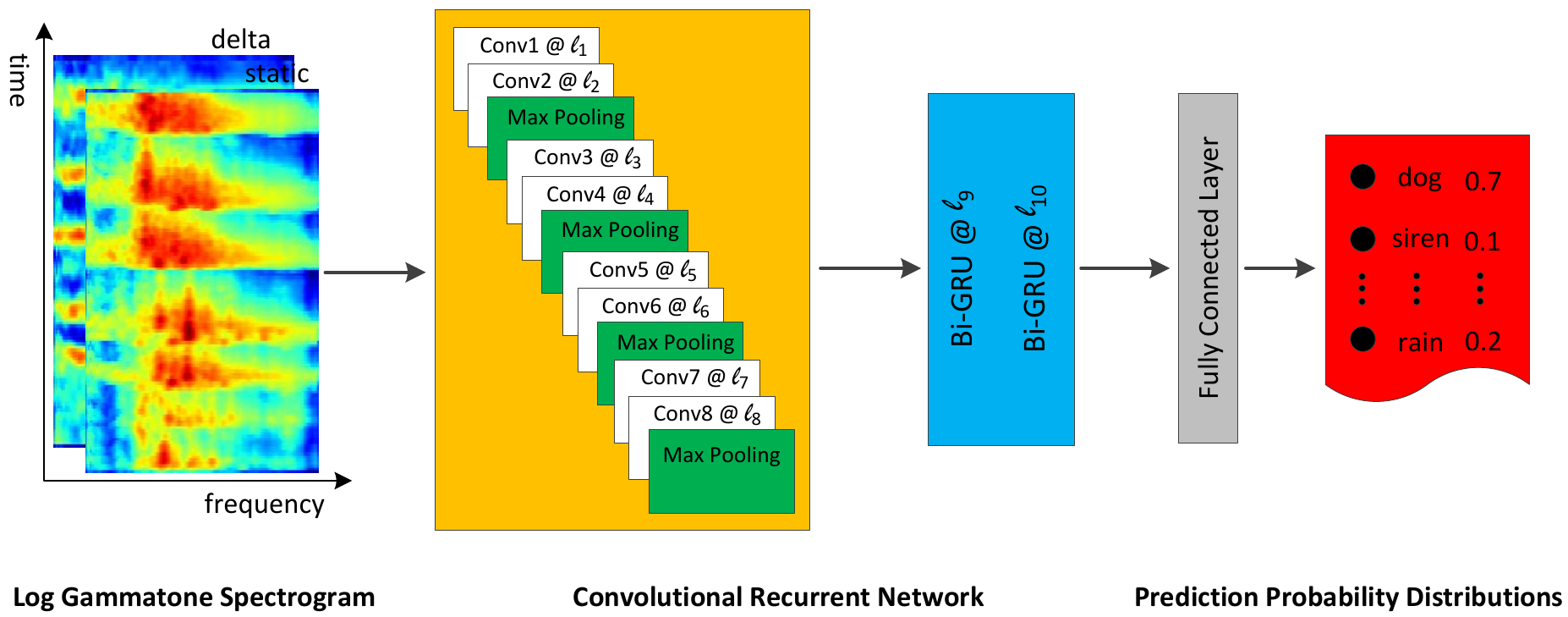}
        \caption{Architecture of convolutional recurrent neural network for environmental sound classification. We extract log gammatone spectrogram and its delta information (Log-GTs) as input to our convolutional recurrent neural network, where the convolutional recurrent neural network consists of eight convolutional layers (\emph{$l_1$}-\emph{$l_8$}) and two bidirectional GRU layers (\emph{$l_9$}-\emph{$l_{10}$}).}
        \label{fig:framework}
\end{figure*}

In this section, we introduce the proposed method for ESC. Firstly, we describe how to generate log gammatone spetrogram (Log-GTs) features from environmental sounds. Then, we introduce the architecture of ACRNN which combines convolutional RNN and a frame-level attention mechanism. The architecture of proposed ACRNN is shown in Figure \ref{fig:framework}. And we will give a detailed description about the architecture of convolutional RNN and the attention mechanism, respectively. Finally, the data augmentation methods are introduced.

\subsection{Feature Extraction and Preprocessing}
Given a signal, we first use short-time Fourier transform (STFT) with hamming window size of 23 ms (1024 samples at 44.1kHz) and 50$\%$ overlap to extract the energy spectrogram. Then, we apply a 128-band gammatone filter bank \cite{valero2012gammatone} to the energy spectrogram and the resulting spectrogram is converted into logarithmic scale. In order to make efficient use of limited data, the spectrogram is split into 128 frames (approximately 1.5s in length) with 50$\%$ overlap. The delta information of the original spectrogram is calculated, which is the first temporal derivative of the static spectrogram. Afterwards, we concatenate the static log gammatone spectrogram and its delta information to a 3-D feature representation $X\in{R^{128\times{128}\times{2}}}$ (Log-GTs) as the input of the network.

\subsection{Architecture of Convolutional RNN}
In this section, we introduce the proposed convolutional RNN which is used to analyze Log-GTs for ESC. We first use CNN to learn high level feature representations from the Log-GTs. Then, the CNN-learned features are fed into bidirectional gated recurrent unit (Bi-GRU) layers which are used to learn the temporal correlation information. Finally, these features are fed into a fully connected layer with a softmax function to output the probability distribution of different classes.
In this paper, the convolutional RNN is comprised of eight convolutional layers (\emph{$l_1$}-\emph{$l_8$}) and two Bi-GRU layers (\emph{$l_9$}-\emph{$l_{10}$}). The architecture and parameters of network are as follows:
\begin{itemize}
    \item  \emph{$l_1$}-\emph{$l_2$}: The first two stacked convolutional layers use 32 filters with a receptive field of (3,5) and stride of (1,1). This is followed by a max-pooling with a (4,3) stride to reduce the dimensions of feature maps. ReLU activation function is used.
    \item  \emph{$l_3$}-\emph{$l_4$}: The next two convolutional layers use 64 filters with a receptive field of (3,1) and stride of (1,1), and is used to learn local patterns along the frequency dimension. This is followed by a max-pooling with a (4,1) stride. ReLU activation function is used.
    \item  \emph{$l_5$}-\emph{$l_6$}: The following pair of convolutional layers uses 128 filters with a receptive field of (1,5) and stride of (1,1), and is used to learn local patterns along the time dimension. This is followed by a max-pooling with a (1,3) stride. ReLU activation function is used.
    \item  \emph{$l_7$}-\emph{$l_8$}: The subsequent two convolutional layers use 256 filters with a receptive field of (3,3) and stride of (1,1) to learn joint time-frequency characteristics. This is followed by a max-pooling of a (2,2) stride. ReLU activation function is used.
    \item  \emph{$l_9$}-\emph{$l_{10}$}: Two Bi-GRU layers with 256 cells are used for temporal summarization, and tanh activation function is used. Dropout with probability of $0.5$ is used for each Bi-GRU layer to avoid overfitting.
\end{itemize}

Batch normalization \cite{ioffe2015batch} is applied to the output of the convolutional layers to speed up training. L2-regularization is applied to the weights of each layer with a coefficient $0.0001$.

\subsection{Frame-level Attention Mechanism}

\begin{figure*}
\centering
        \includegraphics[width=5 in]{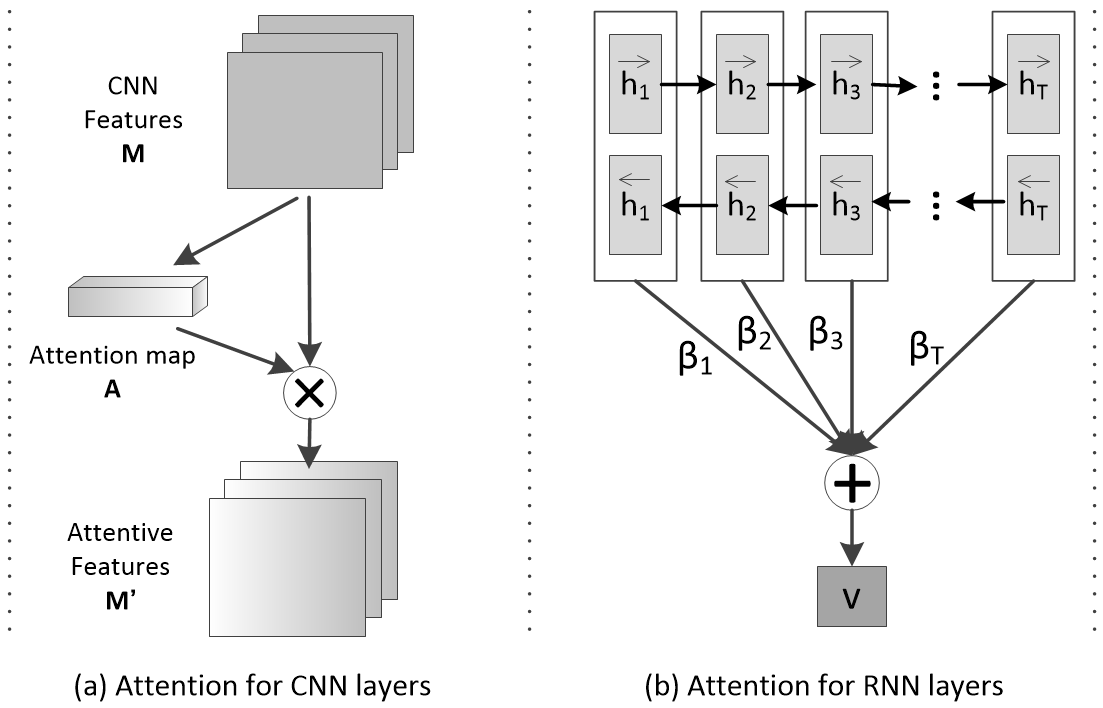}
        \caption{Frame-level attention for (a) CNN layers and (b) RNN layers. For CNN layers, we use frame-level attention to obtain attention map, which is multiplied in frame-wise of CNN features, resulting the attention weighted features. For RNN layers, we utilize frame-level attention to obtain attention weights, which is multiplied in frame-wise of input features. Then, we aggregate these attention weighted representations to form a feature vector, which can be seen as a high-level representation of a sound like \emph{dog bark}.}
        \label{fig:cnn_rnn_attention}
\end{figure*}

Environmental sounds have quite complicated temporal structure and not all frame-level features contribute equally to representations of environmental sounds. Therefore, we apply frame-level attention mechanisms to force model to automatically focus on the meaningful frames and to produce discriminative representations for ESC. In this paper, we investigate frame-level attention mechanisms for CNN layers and RNN layers, respectively.

\subsubsection{Attention for CNN Layers}

As shown in Figure \ref{fig:cnn_rnn_attention}(a), given CNN features $M\in{R^{F\times{T}\times{C}}}$, we first use a 3x3 convolution filter to learn a hidden representation.
This is followed by a average-pool with $(F,1)$ size in order to reduce the frequency dimension to one. Then, we use a scaling function to form a normalized attention map $A\in{R^{1\times{T}\times{1}}}$, which holds the frame-level attention weights for CNN features. The process can be expressed as,

\begin{equation}\label{eqn:cnn_att}
A = \sigma(AvgPool(Conv^{3 \times 3}(M)))
\end{equation}

Where $\sigma$ denotes scaling function. With attention map $A$, the attention weighted CNN features are obtained as,

\begin{equation}\label{eqn:cnn_att}
M'=M\cdot{A}
\end{equation}

The attention is applied by multiplying the attention vector $A$ to each feature vector of $M$ along frequency dimension and channel dimension.

\subsubsection{Attention for RNN Layers:}

As shown in Figure \ref{fig:cnn_rnn_attention}(b), we first feed the Bi-GRU output $h_t=[\overrightarrow{h_t}, \overleftarrow{h_t}]$ through a one-layer MLP to obtain a hidden representation of $h_t$, then we calculate the normalized importance weights $\beta_t$ as,

\begin{equation}\label{eqn:norm}
\beta_t=\frac{exp(W*h_t)}{\sum_{t=1}^T{exp(W*h_t)}}
\end{equation}

Where $W$ is the weight matrix of the MLP. After that, we compute the feature vector $v$ through a weighted sum of the frame-level convolutional RNN feautues based on the weights $\beta_t$ as,

\begin{equation}\label{eqn:sum}
v=\sum_{t=1}^T{\beta_t{h_t}}
\end{equation}

The feature vector $v$ is forwarded into the fully connected layer for final classification.

\subsection{Data Augmentation}
\label{subsect:mixup}

Limited data easily leads model towards overfitting. In this paper, we use time stretch with a factor randomly selected from [0.8, 1.3] and pitch shift with a factor randomly selected from [-3.5, 3.5] to increase raw training data size. In addition, an efficient mixup  \cite{zhang2017mixup} augmentation method is used to construct virtual training data and extend the training distribution. In mixup, the training feature-target pair $(\hat{\mathbf x}, \hat{\mathbf y})$ is generated by mixing two feature-target samples, which is determined by

\begin{equation}\label{eqn:mapping}
\left\{
\begin{aligned}
\hat{\mathbf x}= {\lambda}x_i + (1-\lambda)x_j\\
\hat{\mathbf y}= {\lambda}y_i + (1-\lambda)y_j
\end{aligned}
\right.
\end{equation}

where $x_i$ and $x_j$ are two features randomly selected from the original training Log-GTs, $y_i$ and $y_j$ are their one-hot labels. The mixing factor $\lambda$ is determined by a hyper-parameter $\alpha$ and $\lambda$ $\sim$ Beta($\alpha$, $\alpha$). The training targets used for the mixed samples are produced with the same proportion. The mixing between features and corresponding one-hot labels encourages the model to learn linear interpolated characteristics in-between training examples, which can reduce the amount of undesirable oscillations when we predict samples outside the training examples \cite{zhang2017mixup}. For example, suppose that there is a \emph{dog bark} Log-GTs and a \emph{rain} Log-GTs, whose one-hot labels are $y_1=[1,0,0,0,0,0,0,0,0,0]$ and $y_2=[0,1,0,0,0,0,0,0,0,0]$ respectively.
If we mix the \emph{dog bark} Log-GTs and the \emph{rain} Log-GTs with $\lambda$ equal to 0.7, the training target for the mixed sample should be set as $\hat{\mathbf y}=[0.7,0.3,0,0,0,0,0,0,0,0]$.

\section{Experiments and Results} \label{sect:exp}

\subsection{Experiment Setup}

To evaluate the performance of our proposed method, we carry out experiments on two publicly available datasets: ESC-50 and ESC-10 \cite{piczak2015esc}. ESC-50 is a collection of 2000 environmental recordings containing 50 classes in 5 major categories, including \emph{animals, natural soundscapes and water sounds, human non-speech sounds, interior/domestic sounds}, and \emph{exterior/urban noises}. All audio samples are 5 seconds in duration with a 44.1 kHz sampling rate. ESC-10 is a subset of 10 classes (400 samples) selected from the ESC-50 dataset (\emph{dog bark, rain, sea waves, baby cry, clock tick, person sneeze, helicopter, chainsaw, rooster, fire crackling}).

In this paper, we use a sampling rate of 44.1 kHz for all samples in order to use rich high-frequency information. For training, all models optimize cross-entropy loss using mini-batch stochastic gradient descent with Nesterov momentum of 0.9. Each batch consists of 64 segments randomly selected from the training set without repetition. All models are trained for 300 epochs by beginning with an initial learning rate of 0.01, and then divided the learning rate by 10 every 100 epochs. We initialize the network weights to zero mean Gaussian noise with a standard deviation of 0.05. In the test phase, we evaluate the whole sample prediction with the highest average prediction probability of each segment. Both the training and testing features are normalized by the global mean and stardard deviation of the training set. All models are trained using Keras library with TensorFlow backend on a Nvidia P100 GPU with 12GB memory.

\subsection{Experiment Results}

\begin{figure*}
\centering
        \includegraphics[width=4.6 in]{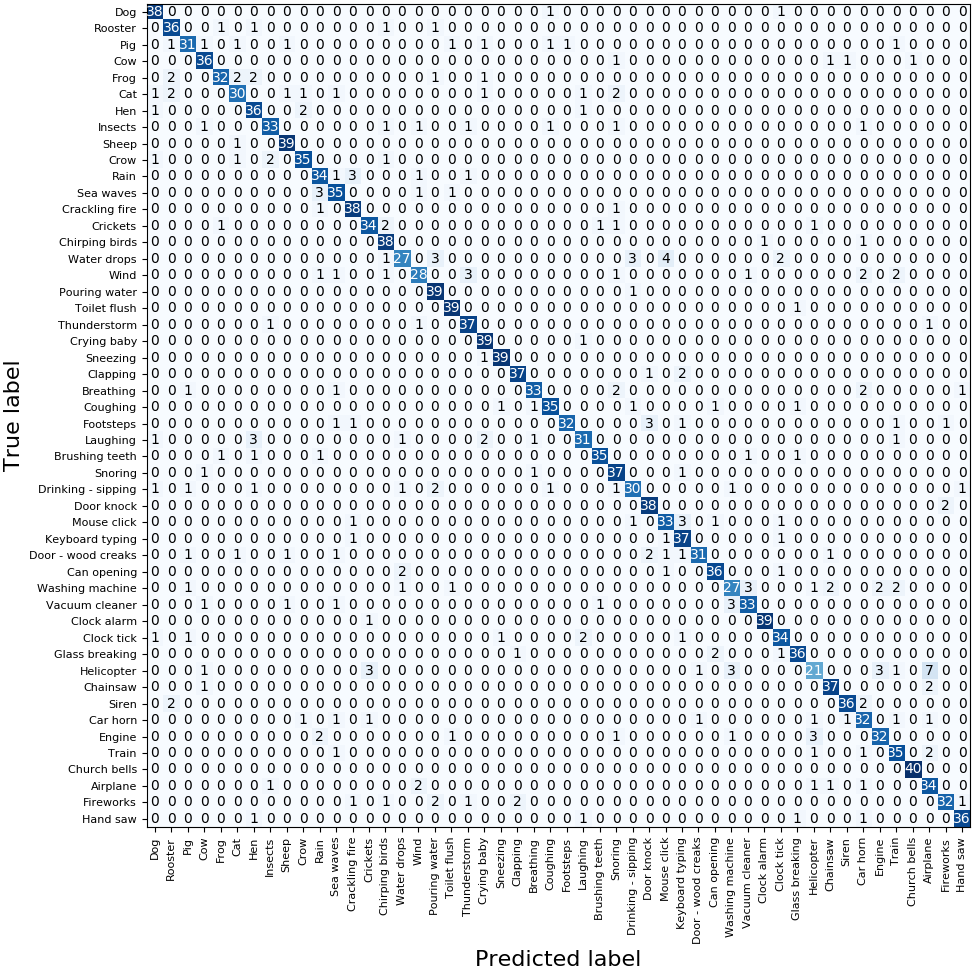}
        \caption{Confusion matrix of ACRNN with an average classification accuracy 86.1$\%$ on ESC-50 dataset.}
        \label{fig:conf_esc50}
\end{figure*}

\begin{table}
\caption{Comparison of ACRNN and existing methods. We perform 5-fold cross validation (CV) by using the official fold settings. The average results of CV are recorded.
\label{tab:res}}
\centering  %
\setlength{\tabcolsep}{5mm}{
\renewcommand\arraystretch{1.2}
\begin{tabular}{lcccccccc}
\hline
\hline
\textbf{Model} &\textbf{ESC-10} &\textbf{ESC-50}\\
\hline
KNN \cite{piczak2015escdata} &66.7\% &32.3\%\\
\hline
SVM \cite{piczak2015escdata} &67.5\% &39.6\%\\
\hline
Random Forest \cite{piczak2015escdata} &72.7\% &44.3\%\\
\hline
AlexNet \cite{boddapati2017classifying} &78.4\% &78.7\%\\
\hline
Google Net \cite{boddapati2017classifying} &63.2\% &67.8\%\\
\hline
PiczakCNN \cite{piczak2015environmental} &80.5\% &64.9\%\\
\hline
SoundNet \cite{aytar2016soundnet} &92.1\% &74.2\%\\
\hline
WaveMsNet \cite{zhu2018learning} &93.7\% &79.1\%\\
\hline
EnvNet-v2 \cite{tokozume2017learning} &91.4\% &84.9\%\\
\hline
ProCNN \cite{li2018ensemble} &92.1\% &82.8\%\\
\hline
Multi-Stream CNN \cite{li2019multi} &93.7\% &83.5\%\\
\hline
ACRNN &\bf{93.7\%} &\bf{86.1\%}\\
\hline
\hline
\end{tabular}}
\end{table}

We compare our model with existing methods reported as PiczakCNN \cite{piczak2015environmental}, SoundNet \cite{aytar2016soundnet}, WaveMsNet \cite{zhu2018learning}, EnvNet-v2 \cite{tokozume2017learning} and Multi-Stream CNN \cite{li2019multi}. According to  \cite{piczak2015environmental}, PiczakCNN consists of two convolutional layers and three fully connected layers. The input features of CNN are generated by combining log mel spectrogram and its delta information. We refer PiczakCNN as a baseline method.

The results are summarized in Table \ref{tab:res}. We see that ACRNN outperforms PiczakCNN and obtains an absolute improvement of 13.2$\%$ and 21.2$\%$ on ESC-10 and ESC-50 datasets, respectively. Then, we compare our model with several state-of-the-art methods: SoundNet8 \cite{aytar2016soundnet}, WaveMsNet \cite{zhu2018learning}, EnvNet-v2 \cite{tokozume2017learning} and Multi-Stream CNN \cite{li2019multi}. We observe that on both ESC-10 and ESC-50 datasets, ACRNN obtains the highest classification accuracy. Note that WaveMsNet \cite{zhu2018learning} and Multi-Stream CNN \cite{li2019multi} achieve the same classification accuracy as ACRNN on ESC-10 but using feature fusion (raw data and spectrogram features), whereas ACRNN only utilizes spectrogram features.

In Figure \ref{fig:conf_esc50}, we provide the confusion matrix generated by ACRNN for ESC-50 dataset. We see that most classes achieve higher accuracy than 80$\%$(32/40). Particularly, \emph{church bells} obtains a 100$\%$ recognition rate. However, we observe that only 52.5$\%$(21/40) \emph{helicopter} samples are correctly recognized with 17.5$\%$(7/40) samples misclassified as \emph{airplane}. We attribute this mistakes to the similar characteristics between the two environmental sounds.

\begin{table}
\caption{Computational complexity of PiczakCNN and proposed ACRNN. "Params" denotes the number of model parameters and "FLOPs" denotes the number of floating point operations.
\label{tab:compute_cost}}
\centering  %
\setlength{\tabcolsep}{2.6mm}{
\renewcommand\arraystretch{1.2}
\begin{tabular}{lcccccccc}
\hline
\hline
\textbf{Model} &\textbf{Params (M)} &\textbf{FLOPs (M)} \\
\hline
PiczakCNN \cite{piczak2015environmental} & 31.53 & 63.27 \\
\hline
convolutional RNN (Ours) & 3.81 & 9.17 \\
\hline
ACRNN (Ours) & 3.81 & 9.18 \\
\hline
\hline
\end{tabular}}
\end{table}

In order to further demonstrate the efficiency of the proposed method, we compare the computational complexity of our proposed approach and PiczakCNN \cite{piczak2015environmental} that is usually referred as a baseline of neural network methods for ESC tasks. We reproduced the network architecture of PiczakCNN according to the description provided by \cite{piczak2015environmental}. Table \ref{tab:compute_cost} shows the number of model parameters and floating point operations (FLOPs) of the proposed method and PiczakCNN. From Table \ref{tab:compute_cost}, we can see that the total parameters and FLOPs of ACRNN are significantly lower than PiczakCNN, while ACRNN outperforms PiczakCNN in terms of classification accuracy. In addition, we also compare the computational complexity of the convolutional RNN (without attention) and ACRNN (with attention at $l_{10}$). The result shows that the model parameters of the two models are almost same with 3.81 million and the FLOPs
of ACRNN is only 0.01 million more than convolutional RNN, which indicates that with the computational complexity almost unchanged, the proposed method can effectively improve the classification accuracy.

\subsection{Effects of attention mechanism}

\begin{table}
\caption{Classification accuracy of proposed convolutional RNN with and without the attention mechanism. 'augment' denotes a combination of time stretch, pitch shift and mixup.
\label{tab:attention}}
\centering  %
\setlength{\tabcolsep}{5mm}{
\renewcommand\arraystretch{1.2}
\begin{tabular}{lcccccccc}
\hline
\hline
\textbf{Model Settings} &\textbf{ESC-10} &\textbf{ESC-50}\\
\hline
convolutional RNN &89.2\% &79.9\%\\
\hline
+ attention &91.7\% &81.3\%\\
\hline
+ augment &93.0\% &84.6\%\\
\hline
+ attention + augment &\textbf{93.7\%} &\textbf{86.1\%}\\
\hline
\hline
\end{tabular}}
\end{table}

To investigate the effect of the attention mechanism, we compare the results of proposed convolutional RNN with and without the attention mechanism. In Table \ref{tab:attention}, the results show that the attention mechanism delivers a significantly improved accuracy even when we use a data augmentation scheme. In addition, the data augmentation boosts an improvement of 2.0$\%$ and 4.8$\%$ on ESC-10 and ESC-50 datasets, respectively.

\subsection{Where and how to apply attention}

\begin{table}
\caption{Classification accuracy of applying the attention mechanism to the output of different layers of the proposed convolutional RNN and using different scaling functions.
\label{tab:cnn_attention}}
\centering  %
\setlength{\tabcolsep}{5mm}{
\renewcommand\arraystretch{1.2}
\begin{tabular}{lcccccccc}
\hline
\hline
\textbf{Model Settings} &\textbf{ESC-10} &\textbf{ESC-50}\\
\hline
no attention &93.0\% &84.6\%\\
\hline
\hline
attention at \emph{$l_2$} (softmax) &93.5\% &85.2\%\\
attention at \emph{$l_2$} (sigmoid) &93.5\% &85.6\%\\
\hline
attention at \emph{$l_4$} (softmax) &92.7\% &83.8\%\\
attention at \emph{$l_4$} (sigmoid) &93.5\% &85.0\%\\
\hline
attention at \emph{$l_6$} (softmax) &92.7\% &84.4\%\\
attention at \emph{$l_6$} (sigmoid) &93.0\% &84.8\%\\
\hline
attention at \emph{$l_8$} (softmax) &92.5\% &84.9\%\\
attention at \emph{$l_8$} (sigmoid) &93.2\% &85.0\%\\
\hline
\hline
attention at \emph{$l_{10}$} &\textbf{93.7\%} &\textbf{86.1\%}\\
\hline
\hline
\end{tabular}}
\end{table}

In this section, we first investigate the classification performance when applying frame-level attention mechanism to the different layers of CNN and RNN. Specifically, we conduct experiments about applying frame-level attention mechanism to $l_2$, $l_4$, $l_6$, $l_8$ and $l_{10}$ layers of our convolutional RNN network. As shown in Table \ref{tab:cnn_attention}, our model obtains the highest classification accuracy and boosts an absolutely improvement of 0.7\% and 1.5\% when applying the attention mechanism at \emph{$l_{10}$} on both ESC-10 and ESC-50 datasets, respectively. For CNN layers, applying the attention mechanism at $l_2$ layer can obtain better classification accuracy than others. In addition, we observe that applying the attention mechanism at lower-layers ($l_2$ and $l_4$) will obtain better performance than applying it at higher-level layers ($l_6$ and $l_8$). We argue that lower-level features usually keep the basic and useful characteristics of environmental sounds and the attention mechanism can help preserve them.

In addition, we evaluate the effect of scaling function on classification accuracy. Specifically, we select softmax and sigmoid function as scaling function and compare their classification accuracy. From Table \ref{tab:cnn_attention} we can see that when we select softmax as scaling function, our model only obtains a slight improvement with the attention mechanism applied at $l_2$ and $l_8$ on ESC-50 dataset and $l_2$ on ESC-10 dataset. When applying attention mechanism at other CNN layers, the classification accuracy decreased. However, when we select sigmoid as scaling function, the proposed frame-level attention can always improve the classification accuracy on ESC-10 and ESC-50 datasets.

\subsection{Visualization of attention}\label{subsect:visul}

\begin{figure*}
\centering
        \includegraphics[width=6.5 in]{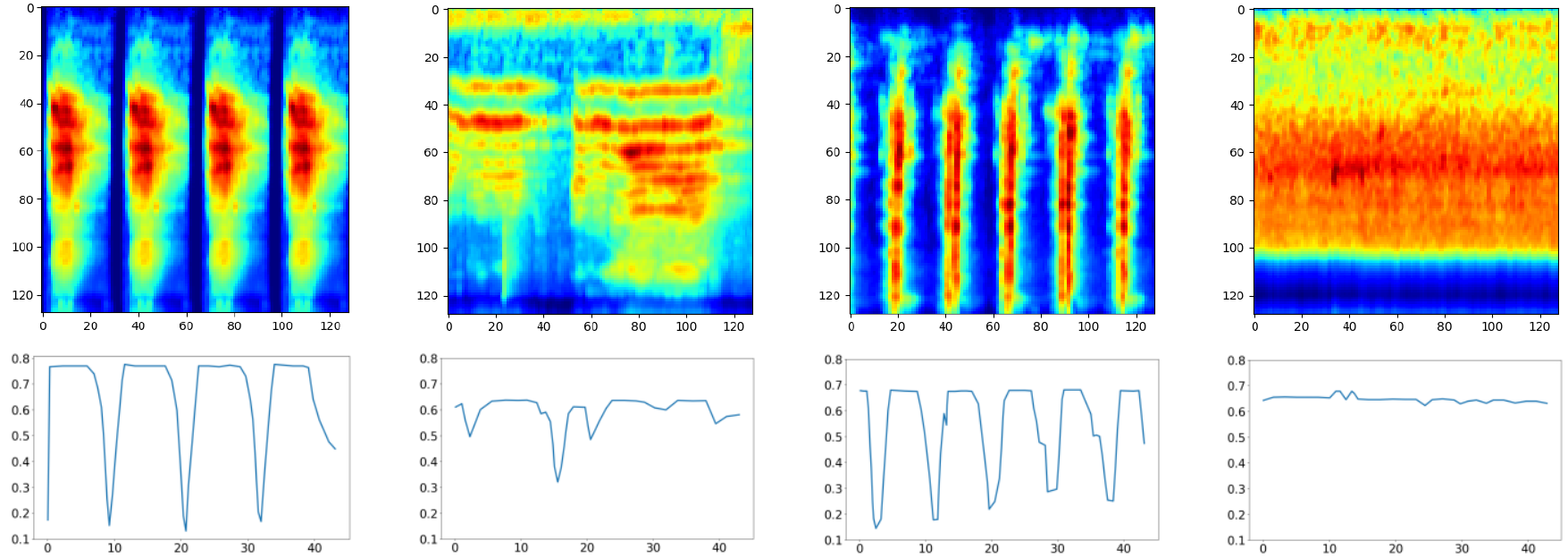}
        \caption{Visualization of frame-level attention results with sigmoid as scaling fucntion at \emph{$l_2$} layer. The first row represents the log gammatone spectrogram of \emph{dog bark}, \emph{crying baby}, \emph{clock tick} and \emph{rain}, the second row indicates the learned frame-level attention weights.}
        \label{fig:visual_att}
\end{figure*}

To have a better understanding how the proposed frame-level attention helps recognize different environmental sounds, we visualize the attention results for different sound classes (\emph{dog bark, crying baby, clock tick} and \emph{rain}). From Figure \ref{fig:visual_att}, we can see that the proposed frame-level attention mechanism is capable of focusing on the important temporal events while reducing the impact of background noise. For example, when our model tries to predict a \emph{dog bark}, our frame-level attention will assign more weights on the semantically relevant frames, while de-weighting the semantically irrelevant or silent ones.

\section{Conclusion} \label{sect:conc}
In this paper, we proposed an attention mechanism based convolutional recurrent neural network (ACRNN) for ESC. We investigated the frame-level attention mechanism for CNN layers and RNN layers. In addition, we compared the computational complexity of our proposed approach and existing work. Experimental results on ESC-10 and ESC-50 datasets demonstrated the effectiveness of the proposed method and achieved state-of-the-art or competitive classification accuracy with low computational complexity. We also compared the classification accuracy when applying different layers, including CNN layers and RNN layers. For CNN layers, we explored the effects of using different attention scaling function on classification accuracy. The experimental results showed that applying attention for RNN layers obtained highest accuracy and sigmoid worked better for generating attention weights than softmax when applying attention at CNN layers. Finally, we visualized the learned attention results to have a better understanding how attention works. While the proposed method achieves the promising results, the robustness to noise of the proposed method is not quantified in this paper. Therefore, in the future work, we will evaluate the robustness to noise of the proposed method with different types of noise with different levels of SNR.

\bibliographystyle{cas-model2-names}
\bibliography{reference}

\begin{thebibliography}{33}
\expandafter\ifx\csname natexlab\endcsname\relax\def\natexlab#1{#1}\fi
\providecommand{\url}[1]{\texttt{#1}}
\providecommand{\href}[2]{#2}
\providecommand{\path}[1]{#1}
\providecommand{\DOIprefix}{doi:}
\providecommand{\ArXivprefix}{arXiv:}
\providecommand{\URLprefix}{URL: }
\providecommand{\Pubmedprefix}{pmid:}
\providecommand{\doi}[1]{\href{http://dx.doi.org/#1}{\path{#1}}}
\providecommand{\Pubmed}[1]{\href{pmid:#1}{\path{#1}}}
\providecommand{\bibinfo}[2]{#2}
\ifx\xfnm\relax \def\xfnm[#1]{\unskip,\space#1}\fi
\bibitem[{Aytar et~al.(2016)Aytar, Vondrick and Torralba}]{aytar2016soundnet}
\bibinfo{author}{Aytar, Y.}, \bibinfo{author}{Vondrick, C.},
  \bibinfo{author}{Torralba, A.}, \bibinfo{year}{2016}.
\newblock \bibinfo{title}{{Soundnet: Learning Sound Representations from
  Unlabeled Video}}, in: \bibinfo{booktitle}{Proc. Int. Conf. Neural Inf.
  Process. Syst.}, pp. \bibinfo{pages}{892--900}.
\bibitem[{Bahdanau et~al.(2014)Bahdanau, Cho and Bengio}]{bahdanau2014neural}
\bibinfo{author}{Bahdanau, D.}, \bibinfo{author}{Cho, K.},
  \bibinfo{author}{Bengio, Y.}, \bibinfo{year}{2014}.
\newblock \bibinfo{title}{{Neural Machine Translation by Jointly Learning to
  Align and Translate}}.
\newblock \bibinfo{journal}{arXiv preprint arXiv:1409.0473} .
\bibitem[{Barchiesi et~al.(2015)Barchiesi, Giannoulis, Stowell and
  Plumbley}]{barchiesi2015acoustic}
\bibinfo{author}{Barchiesi, D.}, \bibinfo{author}{Giannoulis, D.},
  \bibinfo{author}{Stowell, D.}, \bibinfo{author}{Plumbley, M.D.},
  \bibinfo{year}{2015}.
\newblock \bibinfo{title}{{Acoustic Scene Classification: Classifying
  Environments from the Sounds They Produce}}.
\newblock \bibinfo{journal}{IEEE Signal Process. Magazine}
  \bibinfo{volume}{32}, \bibinfo{pages}{16--34}.
\bibitem[{Bisot et~al.(2017)Bisot, Serizel, Essid and
  Richard}]{bisot2017feature}
\bibinfo{author}{Bisot, V.}, \bibinfo{author}{Serizel, R.},
  \bibinfo{author}{Essid, S.}, \bibinfo{author}{Richard, G.},
  \bibinfo{year}{2017}.
\newblock \bibinfo{title}{{Feature Learning with Matrix Factorization Applied
  to Acoustic Scene Classification}}.
\newblock \bibinfo{journal}{IEEE/ACM Trans. Audio, Speech, Language Process.}
  \bibinfo{volume}{25}, \bibinfo{pages}{1216--1229}.
\bibitem[{Boddapati et~al.(2017)Boddapati, Petef, Rasmusson and
  Lundberg}]{boddapati2017classifying}
\bibinfo{author}{Boddapati, V.}, \bibinfo{author}{Petef, A.},
  \bibinfo{author}{Rasmusson, J.}, \bibinfo{author}{Lundberg, L.},
  \bibinfo{year}{2017}.
\newblock \bibinfo{title}{{Classifying Environmental Sounds Using Image
  Recognition Networks}}.
\newblock \bibinfo{journal}{Procedia Computer Science} \bibinfo{volume}{112},
  \bibinfo{pages}{2048--2056}.
\bibitem[{Chorowski et~al.(2015)Chorowski, Bahdanau, Serdyuk, Cho and
  Bengio}]{chorowski2015attention}
\bibinfo{author}{Chorowski, J.K.}, \bibinfo{author}{Bahdanau, D.},
  \bibinfo{author}{Serdyuk, D.}, \bibinfo{author}{Cho, K.},
  \bibinfo{author}{Bengio, Y.}, \bibinfo{year}{2015}.
\newblock \bibinfo{title}{{Attention-based Models for Speech Recognition}}, in:
  \bibinfo{booktitle}{Proc. Int. Conf. Neural Inf. Process. Syst.}, pp.
  \bibinfo{pages}{577--585}.
\bibitem[{Chu et~al.(2009)Chu, Narayanan and Kuo}]{chu2009environmental}
\bibinfo{author}{Chu, S.}, \bibinfo{author}{Narayanan, S.},
  \bibinfo{author}{Kuo, C.C.J.}, \bibinfo{year}{2009}.
\newblock \bibinfo{title}{{Environmental Sound Recognition with Time--Frequency
  Audio Features}}.
\newblock \bibinfo{journal}{IEEE Trans. Audio, Speech, Language Process.}
  \bibinfo{volume}{17}, \bibinfo{pages}{1142--1158}.
\bibitem[{Dai et~al.(2017)Dai, Dai, Qu, Li and Das}]{dai2017very}
\bibinfo{author}{Dai, W.}, \bibinfo{author}{Dai, C.}, \bibinfo{author}{Qu, S.},
  \bibinfo{author}{Li, J.}, \bibinfo{author}{Das, S.}, \bibinfo{year}{2017}.
\newblock \bibinfo{title}{{Very Deep Convolutional Neural Networks for Raw
  Waveforms}}, in: \bibinfo{booktitle}{Proc. Int. Conf. Acoust., Speech, Signal
  Process.}, pp. \bibinfo{pages}{421--425}.
\bibitem[{Dhanalakshmi et~al.(2011)Dhanalakshmi, Palanivel and
  Ramalingam}]{dhanalakshmi2011classification}
\bibinfo{author}{Dhanalakshmi, P.}, \bibinfo{author}{Palanivel, S.},
  \bibinfo{author}{Ramalingam, V.}, \bibinfo{year}{2011}.
\newblock \bibinfo{title}{{Classification of Audio Signals Using AANN and
  GMM}}.
\newblock \bibinfo{journal}{Applied Soft Comput.} \bibinfo{volume}{11},
  \bibinfo{pages}{716--723}.
\bibitem[{Geiger and Helwani(2015)}]{geiger2015improving}
\bibinfo{author}{Geiger, J.T.}, \bibinfo{author}{Helwani, K.},
  \bibinfo{year}{2015}.
\newblock \bibinfo{title}{{Improving Event Detection for Audio Surveillance
  Using Gabor Filterbank Features}}, in: \bibinfo{booktitle}{Proc. Euro. Signal
  Process. Conf.}, pp. \bibinfo{pages}{714--718}.
\bibitem[{Guo et~al.(2017)Guo, Xu, Li and Alwan}]{guo2017attention}
\bibinfo{author}{Guo, J.}, \bibinfo{author}{Xu, N.}, \bibinfo{author}{Li,
  L.J.}, \bibinfo{author}{Alwan, A.}, \bibinfo{year}{2017}.
\newblock \bibinfo{title}{{Attention Based CLDNNs for Short-Duration Acoustic
  Scene Classification}}, in: \bibinfo{booktitle}{Proc. Interspeech}, pp.
  \bibinfo{pages}{469--473}.
\bibitem[{Ioffe and Szegedy(2015)}]{ioffe2015batch}
\bibinfo{author}{Ioffe, S.}, \bibinfo{author}{Szegedy, C.},
  \bibinfo{year}{2015}.
\newblock \bibinfo{title}{{Batch Normalization: Accelerating Deep Network
  Training by Reducing Internal Covariate Shift}}.
\newblock \bibinfo{journal}{arXiv preprint arXiv:1502.03167} .
\bibitem[{Jun and Shengchen(2018)}]{WJ2018ASC}
\bibinfo{author}{Jun, W.}, \bibinfo{author}{Shengchen, L.},
  \bibinfo{year}{2018}.
\newblock \bibinfo{title}{{Self-Attention Mechanism Based System for DCASE2018
  Challenge Task1 and Task4}}.
\newblock \bibinfo{journal}{DCASE2018 Challenge, Tech. Rep.} .
\bibitem[{Li et~al.(2018)Li, Yao, Hu, Liu, Yao and Hu}]{li2018ensemble}
\bibinfo{author}{Li, S.}, \bibinfo{author}{Yao, Y.}, \bibinfo{author}{Hu, J.},
  \bibinfo{author}{Liu, G.}, \bibinfo{author}{Yao, X.}, \bibinfo{author}{Hu,
  J.}, \bibinfo{year}{2018}.
\newblock \bibinfo{title}{{An Ensemble Stacked Convolutional Neural Network
  Model for Environmental Event Sound Recognition}}.
\newblock \bibinfo{journal}{Applied Sciences} \bibinfo{volume}{8},
  \bibinfo{pages}{1152}.
\bibitem[{Li et~al.(2019)Li, Chebiyyam and Kirchhoff}]{li2019multi}
\bibinfo{author}{Li, X.}, \bibinfo{author}{Chebiyyam, V.},
  \bibinfo{author}{Kirchhoff, K.}, \bibinfo{year}{2019}.
\newblock \bibinfo{title}{{Multi-stream Network with Temporal Attention for
  Environmental Sound Classification}}.
\newblock \bibinfo{journal}{arXiv preprint arXiv:1901.08608} .
\bibitem[{Lyon(2010)}]{lyon2010machine}
\bibinfo{author}{Lyon, R.F.}, \bibinfo{year}{2010}.
\newblock \bibinfo{title}{{Machine Hearing: An Emerging Field [Exploratory
  DSP]}}.
\newblock \bibinfo{journal}{IEEE Signal Process. Magazine}
  \bibinfo{volume}{27}, \bibinfo{pages}{131--139}.
\bibitem[{McLoughlin et~al.(2015)McLoughlin, Zhang, Xie, Song and
  Xiao}]{mcloughlin2015robust}
\bibinfo{author}{McLoughlin, I.}, \bibinfo{author}{Zhang, H.},
  \bibinfo{author}{Xie, Z.}, \bibinfo{author}{Song, Y.}, \bibinfo{author}{Xiao,
  W.}, \bibinfo{year}{2015}.
\newblock \bibinfo{title}{{Robust Sound Event Classification Using Deep Neural
  Networks}}.
\newblock \bibinfo{journal}{IEEE/ACM Trans. Audio, Speech, Language Process.}
  \bibinfo{volume}{23}, \bibinfo{pages}{540--552}.
\bibitem[{Mesaros et~al.(2018)Mesaros, Heittola, Benetos, Foster, Lagrange,
  Virtanen and Plumbley}]{mesaros2018detection}
\bibinfo{author}{Mesaros, A.}, \bibinfo{author}{Heittola, T.},
  \bibinfo{author}{Benetos, E.}, \bibinfo{author}{Foster, P.},
  \bibinfo{author}{Lagrange, M.}, \bibinfo{author}{Virtanen, T.},
  \bibinfo{author}{Plumbley, M.D.}, \bibinfo{year}{2018}.
\newblock \bibinfo{title}{{Detection and Classification of Acoustic Scenes and
  Events: Outcome of the DCASE 2016 Challenge}}.
\newblock \bibinfo{journal}{IEEE/ACM Trans. Audio Speech Lang. Process.}
  \bibinfo{volume}{26}, \bibinfo{pages}{379--393}.
\bibitem[{Piczak(2015a)}]{piczak2015environmental}
\bibinfo{author}{Piczak, K.J.}, \bibinfo{year}{2015}a.
\newblock \bibinfo{title}{{Environmental Sound Classification with
  Convolutional Neural Networks}}, in: \bibinfo{booktitle}{Proc. 25th Int.
  Workshop Mach. Learning Signal Process.}, pp. \bibinfo{pages}{1--6}.
\bibitem[{Piczak(2015b)}]{piczak2015esc}
\bibinfo{author}{Piczak, K.J.}, \bibinfo{year}{2015}b.
\newblock \bibinfo{title}{{ESC: Dataset for Environmental Sound
  Classification}}, in: \bibinfo{booktitle}{Proc. 23rd ACM Int. Conf.
  Multimedia}, pp. \bibinfo{pages}{1015--1018}.
\bibitem[{Piczak(2015c)}]{piczak2015escdata}
\bibinfo{author}{Piczak, K.J.}, \bibinfo{year}{2015}c.
\newblock \bibinfo{title}{{ESC: Dataset for Environmental Sound
  Classification}}, in: \bibinfo{booktitle}{Proc. Int. Conf. Multimedia}, pp.
  \bibinfo{pages}{1015--1018}.
\bibitem[{Radhakrishnan et~al.(2005)Radhakrishnan, Divakaran and
  Smaragdis}]{radhakrishnan2005audio}
\bibinfo{author}{Radhakrishnan, R.}, \bibinfo{author}{Divakaran, A.},
  \bibinfo{author}{Smaragdis, A.}, \bibinfo{year}{2005}.
\newblock \bibinfo{title}{{Audio Analysis for Surveillance Applications}}, in:
  \bibinfo{booktitle}{Proc. IEEE Workshop Appl. Signal Process. Audio Acoust.},
  pp. \bibinfo{pages}{158--161}.
\bibitem[{Ren and et. al.(2018)}]{ZR2018ASC}
\bibinfo{author}{Ren, Z.}, \bibinfo{author}{et. al.}, \bibinfo{year}{2018}.
\newblock \bibinfo{title}{{Attention-based Convolutional Neural Networks for
  Acoustic Scene Classification}}.
\newblock \bibinfo{journal}{DCASE2018 Challenge, Tech. Rep.} .
\bibitem[{Sankaran et~al.(2016)Sankaran, Mi, Al-Onaizan and
  Ittycheriah}]{sankaran2016temporal}
\bibinfo{author}{Sankaran, B.}, \bibinfo{author}{Mi, H.},
  \bibinfo{author}{Al-Onaizan, Y.}, \bibinfo{author}{Ittycheriah, A.},
  \bibinfo{year}{2016}.
\newblock \bibinfo{title}{{Temporal Attention Model for Neural Machine
  Translation}}.
\newblock \bibinfo{journal}{arXiv preprint arXiv:1608.02927} .
\bibitem[{Tokozume et~al.(2017)Tokozume, Ushiku and
  Harada}]{tokozume2017learning}
\bibinfo{author}{Tokozume, Y.}, \bibinfo{author}{Ushiku, Y.},
  \bibinfo{author}{Harada, T.}, \bibinfo{year}{2017}.
\newblock \bibinfo{title}{{Learning from Between-Class Examples for Deep Sound
  Recognition}}.
\newblock \bibinfo{journal}{arXiv preprint arXiv:1711.10282} .
\bibitem[{Vacher et~al.(2007)Vacher, Serignat and Chaillol}]{vacher2007sound}
\bibinfo{author}{Vacher, M.}, \bibinfo{author}{Serignat, J.F.},
  \bibinfo{author}{Chaillol, S.}, \bibinfo{year}{2007}.
\newblock \bibinfo{title}{{Sound Classification in A Smart Room Environment: An
  Approach Using GMM and HMM Methods}}, in: \bibinfo{booktitle}{Proc. 4th IEEE
  Conf. Speech Technique, Human-Computer Dialogue}, pp.
  \bibinfo{pages}{135--146}.
\bibitem[{Valero and Alias(2012)}]{valero2012gammatone}
\bibinfo{author}{Valero, X.}, \bibinfo{author}{Alias, F.},
  \bibinfo{year}{2012}.
\newblock \bibinfo{title}{{Gammatone Cepstral Coefficients: Biologically
  Inspired Features for Non-Speech Audio Classification}}.
\newblock \bibinfo{journal}{IEEE Trans. Multimedia} \bibinfo{volume}{14},
  \bibinfo{pages}{1684--1689}.
\bibitem[{Vu and Wang(2016)}]{vu2016acoustic}
\bibinfo{author}{Vu, T.H.}, \bibinfo{author}{Wang, J.C.}, \bibinfo{year}{2016}.
\newblock \bibinfo{title}{{Acoustic Scene and Event Recognition Using Recurrent
  Neural Networks}}.
\newblock \bibinfo{journal}{DCASE2016 Challenge, Tech. Rep.} .
\bibitem[{Yang et~al.(2016)Yang, Yang, Dyer, He, Smola and
  Hovy}]{yang2016hierarchical}
\bibinfo{author}{Yang, Z.}, \bibinfo{author}{Yang, D.}, \bibinfo{author}{Dyer,
  C.}, \bibinfo{author}{He, X.}, \bibinfo{author}{Smola, A.},
  \bibinfo{author}{Hovy, E.}, \bibinfo{year}{2016}.
\newblock \bibinfo{title}{{Hierarchical Attention Networks for Document
  Classification}}, in: \bibinfo{booktitle}{Proc. NAACL-HLT}, pp.
  \bibinfo{pages}{1480--1489}.
\bibitem[{Zhang et~al.(2017)Zhang, Cisse, Dauphin and
  Lopez-Paz}]{zhang2017mixup}
\bibinfo{author}{Zhang, H.}, \bibinfo{author}{Cisse, M.},
  \bibinfo{author}{Dauphin, Y.N.}, \bibinfo{author}{Lopez-Paz, D.},
  \bibinfo{year}{2017}.
\newblock \bibinfo{title}{{Mixup: Beyond Empirical Risk Minimization}}.
\newblock \bibinfo{journal}{arXiv preprint arXiv:1710.09412} .
\bibitem[{Zhang et~al.(2018)Zhang, Xu, Cao and Zhang}]{zhang2018deep}
\bibinfo{author}{Zhang, Z.}, \bibinfo{author}{Xu, S.}, \bibinfo{author}{Cao,
  S.}, \bibinfo{author}{Zhang, S.}, \bibinfo{year}{2018}.
\newblock \bibinfo{title}{{Deep Convolutional Neural Network with Mixup for
  Environmental Sound Classification}}, in: \bibinfo{booktitle}{Proc. Chinese
  Conf. Pattern Recognit. Comput. Vision}, \bibinfo{organization}{Springer}.
  pp. \bibinfo{pages}{356--367}.
\bibitem[{Zhang et~al.(2019)Zhang, Xu, Qiao, Zhang and
  Cao}]{zhang2019attention}
\bibinfo{author}{Zhang, Z.}, \bibinfo{author}{Xu, S.}, \bibinfo{author}{Qiao,
  T.}, \bibinfo{author}{Zhang, S.}, \bibinfo{author}{Cao, S.},
  \bibinfo{year}{2019}.
\newblock \bibinfo{title}{{Attention Based Convolutional Recurrent Neural
  Network for Environmental Sound Classification}}, in:
  \bibinfo{booktitle}{Proc. Chinese Conf. Pattern Recognit. Comput. Vision},
  \bibinfo{organization}{Springer}. pp. \bibinfo{pages}{261--271}.
\bibitem[{Zhu et~al.(2018)Zhu, Wang, Liu, Lei, Lu and Peng}]{zhu2018learning}
\bibinfo{author}{Zhu, B.}, \bibinfo{author}{Wang, C.}, \bibinfo{author}{Liu,
  F.}, \bibinfo{author}{Lei, J.}, \bibinfo{author}{Lu, Z.},
  \bibinfo{author}{Peng, Y.}, \bibinfo{year}{2018}.
\newblock \bibinfo{title}{{Learning Environmental Sounds with Multi-scale
  Convolutional Neural Network}}.
\newblock \bibinfo{journal}{arXiv preprint arXiv:1803.10219} .

\end{thebibliography}

\end{document}